\documentclass[aps,graphicx,twocolumn]{revtex4}
\usepackage{amsmath}
\usepackage{amscd}
\usepackage{graphicx}
\usepackage{multirow}
\usepackage{color}
\begin{document}

\title{Analysis of control power in controlled remote state preparation schemes}
\author{ Xi-Han Li$^{1,2}$\footnote{
Email address: xihanlicqu@gmail.com}, Shohini Ghose$^{2,3}$}
\address{$^1$ Department of Physics, Chongqing University,
Chongqing, China \\$^2$Department of Physics and Computer Science, Wilfrid Laurier University, Waterloo, Canada\\
$^3$ Institute for Quantum Computing, University of Waterloo, Canada}

\date{\today }
\begin{abstract}
We quantify and analyze the controller's power in controlled remote state preparation schemes. Our analysis provides a lower bound on the control power required for controlled remote preparation of arbitrary $D$-dimensional states. We evaluate several existing controlled remote state preparation protocols and show that some proposed non-maximally entangled  channels are not suitable for perfect controlled remote preparation of arbitrary quantum states from the controller's point of view. We find that for remotely preparing $D$-dimensional states, the entropy of each controller should be no less than $\log_2 D$ bits. Our criteria are not only useful for evaluating controlled remote state preparation schemes but can also be used for other controlled quantum communication schemes.
\end{abstract}
\maketitle

\section{Introduction}
Remote state preparation (RSP) is an effective method for transmitting quantum states with the help of pre-shared entanglement and classical communication, but without the need to physically send the states to the receiver \cite{RSP_lo}. The protocol is different from quantum teleportation \cite{tele} in that the sender has classical knowledge about the parameters of a known state to be transmitted. Using this classical information, the sender can perform measurements on his/her own portion of an entangled system. Based on the sender's measurement results, the receiver can recover the desired quantum state with appropriate unitary operations on his/her part of the entangled system. RSP requires less classical communication than quantum teleportation. Moreover, the two main resources in quantum communication - quantum entanglement and classical information -  can be traded off against each other in RSP, while in quantum teleportation they cannot \cite{RSP_Bennet}. Due to its interesting characteristics, RSP has attracted much attention in the past years  \cite{low_entanglement_RSP,optimal_RSP,generalized_RSP,oblivious_RSP,faithful_RSP,cv_RSP} and several RSP experiments have been reported \cite{RSPE1,RSPE2,RSPE3,RSPE4,RSPE5,RSPE6,RSPE7}.

In conventional RSP protocols, there is one sender and one receiver. In order to satisfy different communication requirements, two variants of RSP were proposed -  joint remote state preparation (JRSP) \cite{JRSP_I_GC1,JRSP_I_GC2} and controlled remote state preparation (CRSP) \cite{MCRSP_I_NE_EPR,MCRSP_II_2GC3,CRSP_II_E+G(C),CRSP_II_GC+WC1,CRSP_II_GC+WC2,
CRSP_II_III_B,CRSP_equ,CRSP_I_A4,CRSP_qudit1,CRSP_qudit2,CRSP_partial2}. In JRSP, several senders share knowledge of the state, with each sender only having partial information about the state. In CRSP, a controller is introduced, without whose permission the state cannot be faithfully prepared. The combination of these two kinds of RSP protocols is called controlled joint remote state preparation (CJRSP)\cite{CJRSP_II_E,CJRSP_II_CC8,CJRSP_I,CRSP_partial1}, and can involve multiple senders and controllers. In this letter, we focus on RSP schemes involving one or more controllers. We quantitatively analyze the requirement that the state preparation should only be executed with the participation of the controller/controllers. We refer to these kinds of schemes as controlled-RSP schemes in this article, and include both CRSP and CJRSP schemes.

In 2009, CRSP was first introduced in a scheme to remotely prepare a single-qubit state via the control of many agents \cite{MCRSP_I_NE_EPR}. Later, CRSP of a two-qubit state via a Greenberger-Horne-Zeilinger (GHZ)-class state was also proposed \cite{MCRSP_II_2GC3}. Since then, many controlled-RSP schemes were proposed that exploited different quantum channels such as an asymmetric channel composed of a GHZ-class (GC) state and a W-Class state \cite{CRSP_II_GC+WC1,CRSP_II_GC+WC2}, the Brown state \cite{CRSP_II_III_B}, the Affleck-Kennedy-Lieb-Tasaki (AKLT) state \cite{CRSP_I_A4} and so on.
In addition to CRSP of a two-level state, CRSP schemes for higher-dimensional systems were also investigated \cite{CRSP_qudit1,CRSP_qudit2}. Recently, a CRSP scheme using a partially entangled quantum channel was proposed in which neither auxiliary states nor two-qubit unitary transformations were required \cite{CRSP_partial2}. The success probability of the scheme can be unit for restricted classes of states. A perfect CJRSP scheme that could be implemented deterministically using a partially entangled quantum channel was also proposed~\cite{CRSP_partial1}. The success probabilities of these two schemes are independent of the degree of entanglement (DOE) of the quantum channels, which is surprising and interesting. Our current work sheds light on these counterintuitive results by taking the authority of the controller into account in a quantitative way.

Generally, a quantum communication protocol is evaluated on the basis of security, efficiency and success probability. However, in certain situations, additional factors should be taken into account to perform a comprehensive assessment. In the case of controlled communication protocols, this additional factor is the controller's power. Since the controller plays a key role, the degree to which he/she has control over the protocol must be quantified when assessing the scheme. Based on this idea, we recently proposed a quantitative measure to evaluate perfect controlled teleportation schemes from the controller's point of view \cite{CP1,CP2}.  We identified a lower bound on the control power required for any controlled teleportation scheme. Our analysis showed that in controlled teleportation schemes, the control power needs to be considered in addition to the success probability \cite{CP1} and some existing controlled teleportation schemes are unsuitable for teleporting arbitrary quantum states according to our criterion \cite{CP2}.

In this article, we discuss the controller's power in controlled-RSP protocols and extend the idea of control power to general controlled quantum communication schemes. We present several new results. We first review the definition of control power and describe it in the context of controlled-RSP for qubits. Our measure of control power can be applied to CRSP schemes that succeed only probabilistically. Then we generalize our definition to include controlled-RSP of arbitrary-dimensional systems. Using our measure of control power we evaluate several existing controlled-RSP schemes. We show that previously proposed partially entangled channels \cite{CRSP_partial2,CRSP_partial1} are not suitable for controlled-RSP of arbitrary quantum states. Moreover, we point out that several other controlled-RSP schemes are unsuitable for the task from the controller's point of view. However, some partially entangled channels combined with ancillary states and measurements can be used for RSP  while preserving the controller's power.
Finally, we identify a minimum requirement on the controller's entropy for any controlled quantum communication scheme to successfully preserve the controller's authority.

\section{Controller's power in controlled-RSP}
In Ref. \cite{CP1,CP2}, we discussed the controller's power in the controlled teleportation of both single-qubit states and multi-qubit states. However, the situation was restricted to perfect controlled teleportation in which the success probability should be 100\% in principle. Actually, most of the existing protocols for RSP of arbitrary quantum states succeed probabilistically, even with the use of a maximally entangled channel. Generally, there are two main reasons for the loss of success probability.
The first reason is the impossibility of a universal NOT operation \cite{RSP_pati}, whose success can be judged by the sender's measurement results. The other  reason is the lack of sufficient degree of entanglement of the quantum channel. In this case, the sender can utilize positive-operator valued measures (POVM), or either the sender or the receiver can introduce an additional qubit to help modify the quantum channel to a maximally entangled one in a probabilistic way. The success of the protocol depends on the POVM results or the measurement of the auxiliary state.  Whatever the approach, we can discuss the controller's power only based on successful events. That is, we postpone the controller's actions to the last step after all other parties' (senders/receivers) doable operations and measurements. All the parties' operations commute with each other. Therefore the modification of the order will not influence the scheme at all. Then the nonconditioned fidelity (NCF) of the receiver's state can be calculated based on the state made up of the controller's and receiver's particles after all other parties have performed their operations. In this case, the unsuccessful events are always excluded before this last step.

We can now describe the controller's power in controlled-RSP schemes based on the definition in Ref.\cite{CP1,CP2}. We begin with the simplest situation. Suppose there are three parties, the sender Alice, the receiver Bob and the controller Charlie. The state Alice wants to remotely prepare at Bob's site is $\vert \varphi\rangle$. The information about the state is only known to Alice herself. The entangled system shared between the three parties in advance is $\vert \Theta\rangle_{ABC}$. The subscripts $A$, $B$ and $C$ represent the particle/particles belonging to Alice, Bob and Charlie, respectively. Firstly, the sender Alice performs a measurement on her own particle $A$. The measurement basis is chosen based on her classical knowledge of $\vert \varphi\rangle$. Then the system collapses to $\vert \psi\rangle_{BC}$. Based on Alice's measurement result, Bob knows whether he can get the desired state by performing unitary operations on his particle $B$ or not. If the answer is yes, after appropriate operations performed by Bob, the state changes to $\vert \psi'\rangle_{BC}$. If the controller Charlie doesn't collaborate, then Bob has a mixed state,\begin{eqnarray}
  \rho_B=tr_C(\vert \psi'\rangle_{BC}\langle \psi'\vert).
\end{eqnarray}
The NCF of Bob's state is \cite{CP1,CP2}
\begin{eqnarray}
  f=\langle \varphi\vert \rho_B\vert \varphi\rangle.
\end{eqnarray}
This is the fidelity of the RSP scheme without the help/permission of the controller and it depends on the parameters of the state to be prepared. The average NCF $\bar{f}$ can be calculated by averaging over all possible input states. The control power is defined to be the difference between the NCF and the conditioned fidelity (100\%) when the controller Charlie participates,
\begin{eqnarray}
  P=1-\bar{f}.
\end{eqnarray}
To ensure the controller's authority, the NCF should be as low as possible. For an $N$-qubit state, the classical limit (CL) is
\begin{eqnarray}
  F_{CL}=\frac{2}{1+2^N},
\end{eqnarray}
which is the best fidelity that can be achieved using only classical correlations \cite{cl1,cl2,cl3}. Therefore, the controller's power $P$ should be \cite{CP2}
\begin{eqnarray}
  P\geq\frac{2^N-1}{2^N+1}.
\end{eqnarray}
 This limit is a statement that without the controller's help, the fidelity of the scheme should be no better than the best fidelity achievable via a classical channel.

 This measure can be used in  controlled-RSP schemes with more than one sender and controller and even in the cases where additional states are required. Each controller's power is calculated by letting all other parties perform their operations first and retaining the successful cases. Then the NCF can be computed from the state composed of this controller's and the receiver's particles.

The control power can also be discussed in CRSP schemes of high-dimensional systems. Suppose the dimension of each particle is $d$, and the state is composed of $N$ particles. The total dimension of the system is $D=d^N$. The classical limit for this system is \cite{cl1,cl2,cl3}
\begin{eqnarray}
  F_{CL}=\frac{2}{1+D}=\frac{2}{1+d^N}.
\end{eqnarray}
Therefore the control power should satisfy the following requirement to ensure the controller's authority.
\begin{eqnarray}
  P\geq \frac{D-1}{D+1}. \label{cp}
\end{eqnarray}

We can now use our measure of control power to analyze existing CRSP schemes.

\section{Revisiting CRSP schemes via non-maximally entangled channels}
We first revisit controlled-RSP schemes for preparing a single qubit state that use partially entangled channels.
The two non-maximally entangled channels in Ref.\cite{CRSP_partial2,CRSP_partial1} are
\begin{eqnarray}
  \vert \Theta_{MS_3}\rangle_{ABC}&=&\frac{1}{\sqrt{2}}(\vert 000\rangle+c\vert111\rangle+d\vert 110\rangle),\\
  \vert \Theta_{MS_4}\rangle_{A_1A_2BC}&=&\frac{1}{\sqrt{2}}(\vert 0000\rangle-c\vert1111\rangle+d\vert 1110\rangle).
\end{eqnarray}
 The parameters are both real and  $ c^2+ d^2=1$. The subscript $MS$ refers to a maximal slice state \cite{ms}. The subscripts $A$, $B$ and $C$ denote particles belonging to the three parties Alice (sender), Bob (receiver)and Charlie (controller), respectively. In the second scheme, there are two senders $A_1$ and $A_2$, each holding one particle. The discussion can be made simple and clear by applying local unitaries to rotate these two states to
\begin{eqnarray}
  &&\vert \Theta_{MS_3}\rangle_{ABC}=a\vert 0\rangle_C\vert\Phi^+\rangle_{AB}+b\vert 1\rangle_C\vert \Phi^-\rangle_{AB},\label{p1}\\
  &&\vert \Theta_{MS_4}\rangle_{A_1A_2BC}=a\vert 0\rangle_C\vert \Phi_3^+\rangle_{A_1A_2B}+b\vert 1\rangle_C\vert\Phi_3^-\rangle_{A_1A_2B}.\nonumber\\
\end{eqnarray}
Here we suppose $a>b$. $\vert \Phi^{\pm}\rangle$ represents the Bell state and $\vert \Phi_3^{\pm}\rangle$ denotes the three-qubit GHZ state:
\begin{eqnarray}
\vert \Phi^\pm\rangle&=&\frac{1}{\sqrt{2}}(\vert 00\rangle\pm \vert 11\rangle),\\
\vert \Phi_3^\pm\rangle&=&\frac{1}{\sqrt{2}}(\vert 000\rangle\pm \vert 111\rangle).
\end{eqnarray}
From the expression above, it is simple to see that after the controller measures his particle $C$ in the basis $\vert 0\rangle/\vert 1\rangle$, the sender and the receiver will always be left with a maximally entangled channel which can be used to complete perfect RSP. Apparently, these two non-maximally entangled channels can implement controlled-RSP just as well as maximally entangled ones and the success probabilities can be 100\% , regardless of  the degree of entanglement of the quantum channels. However, this surprising result is not the whole story. We show here that the lack of degree of entanglement will actually affect the controller's power. After the sender's/senders' measurements, if the controller does not participate, then the receiver has a mixed state after tracing over the controller's qubit,
\begin{eqnarray}
  \rho_B=a^2 \vert\varphi\rangle\langle\varphi\vert+b^2\sigma_z\vert \varphi\rangle\langle\varphi\vert\sigma_z^{\dagger}.
\end{eqnarray}
Here $\sigma_z$ is one of the three Pauli operators whose effect is $\vert j\rangle\rightarrow (-1)^j\vert j\rangle$ $(j=0,1)$.
Then the average NCF can be calculated to be
\begin{eqnarray}
  \bar{f}=a^2+\frac{b^2}{3}.
\end{eqnarray}
The controller's power is thus
\begin{eqnarray}
  P=1-\bar{f}=\frac{2b^2}{3}.
\end{eqnarray}
Since the classical limit on the fidelity for RSP of a single-qubit state is 2/3 \cite{2/3,2/3+}, sufficient control power in these two schemes can be obtained only when $b^2=1/2$. This corresponds to the case where the quantum channel is maximally entangled. For all channels where $b^2 < 1/2$,  the control power is always less than $1/3$ ($D=2$ in Eq.(\ref{cp})), which is below the lower bound for an acceptable controlled-RSP scheme. In other words, the non-maximally entangled quantum channels are unsuitable for controlled-RSP of an arbitrary quantum state in these proposed schemes; the controller's power is insufficient due to the lack of degree of entanglement in the quantum channels. The channels may however be useful for controlled RSP of a restricted set of states as in the case of controlled teleportation \cite{CP1}. Also, the channel can be transformed to a maximally entangled one with the help of additional qubits and measurements, and can then be used for controlled-RSP.

\section{Revisiting general CRSP schemes from the controller's viewpoint}
We next analyze several existing controlled-RSP schemes from the controller's viewpoint. The first protocol we discuss here is the single party controlled CRSP scheme for preparing $N$-qubit equatorial states \cite{CRSP_equ}. In this scheme, the quantum channel is composed of $N-1$ Bell states and one GHZ state. The controller only possesses one particle. After the sender's measurement and corresponding operations, the reduced density matrix of the receiver's state is
\begin{eqnarray}
  \rho_{B}=\frac{1}{2}\vert \varphi\rangle\langle\varphi\vert +\frac{1}{2}\vert \phi\rangle\langle\phi\vert. \label{one}
\end{eqnarray}
Here $\vert \phi\rangle$ is a state related to the desired one $\vert \varphi\rangle$ by a unitary operation.
Therefore the average NCF of receiver's state is
\begin{eqnarray}
  \bar{f}\geq1/2.
\end{eqnarray}
The control power is less than 1/2, which does not meet the criterion for acceptable controlled-RSP of $N$-qubit states ($N$ $>$1). Only when $N$=1, the control power is acceptable where the quantum channel is a GHZ state. The same conclusion can be reached for the scheme in Ref.\cite{CRSP_II_III_B}. For preparing an arbitrary two-qubit state $\vert \varphi\rangle=\alpha\vert 00\rangle+\beta\vert 01\rangle+\gamma\vert 10\rangle+\delta\vert 11\rangle$, the quantum channel is a five-qubit genuine entangled Brown state
\begin{eqnarray}
  \vert \Theta_{Br}\rangle&=&\frac{1}{2}(\vert 001\rangle\vert \Phi^-\rangle+\vert 010\rangle\vert \Psi^-\rangle\nonumber\\&&+\vert 100\rangle\vert \Phi^+\rangle+\vert 111\rangle\vert \Psi^+\rangle)_{A_1A_2CB_1B_2}.
\end{eqnarray}
The subscript denotes the particles held by each of the three parties. The authors  also discuss a quantum channel composed of one Brown state and one Bell state to prepare a three-qubit arbitary quantum state. Notice that the controller only controls one particle. The density matrix of the receiver without the controller's help is the same as Eq.(\ref{one}). The NCF is always larger than the corresponding classical limit, indicating insufficient control power.

In the previous section we showed that controlled RSP schemes using partially entangled channels cannot achieve the minimum required control power. However, additional resources such as auxiliary states and measurements can improve the situation. Let us revisit a general CJRSP scheme in which there are two senders, $M$ controllers and one receiver \cite{CJRSP_I}. The quantum channel is composed of $N$ ($M$+3)-particle partially entangle GHZ (PGHZ) states for CJRSP of an arbitrary $N$-qubit quantum state,
\begin{eqnarray}
  \vert \Theta_{PGHZ}\rangle=(a\vert 00...0\rangle+b\vert 11...1\rangle)_{A_1A_2C_1C_2...C_MB}.
\end{eqnarray}
In this protocol, the success probability depends on the parameters of the quantum channel, i.e., the degree of entanglement of the quantum channel. An auxiliary particle was introduced to adjust the coefficients. We calculate the control power conditioned on the successful situations. If  one of the controllers does not participate, the receiver's state will be
\begin{eqnarray}
  \rho_B=\frac{1}{2^N}\vert \varphi\rangle\langle\varphi\vert +\frac{1}{2^N}\sum^{2^N-1}_{j=1}\vert \phi_j\rangle\langle\phi_j\vert.
\end{eqnarray}
The control power is $P=\frac{2^N-1}{2^N+1}$, which is equal to the lower bound on the required control power for  preparing $N$-qubit states. This means that this partially entangled channel combined with an auxiliary state and extra operations and measurements can ensure enough control power. In essence, the partially entangled state is transformed into a maximally entangled one via these extra resources in a probabilistic way.

We now analyze the CRSP of qudit states \cite{CRSP_qudit1,CRSP_qudit2}. In the first scheme, there are $M$ controllers and hence the $(M+2)$-particle generalized GHZ-class (GGC) state  was used as the quantum channel.
\begin{eqnarray}
  \vert \Theta_{GGC}\rangle=a_0\prod^{M+1}_{m=0}\vert 0\rangle_{A_m}+...+a_{d-1}\prod^{M+1}_{m'=0}\vert d-1\rangle_{A_{m'}}.
\end{eqnarray}
Here the particle $A_0$ belongs to the sender Alice, $A_{M+1}$ belongs to the receiver while the rest are held by the $M$ controllers.
The state to be prepared is an arbitrary single-qudit state
\begin{eqnarray}
  \vert \varphi\rangle=\sum_{j=0}^{d-1}\beta_j\vert j\rangle.
\end{eqnarray}
Firstly, the sender performs a POVM measurement on her particle $A_0$. Then the sender and the $M$ controllers measure their own particles in the generalized $X$ basis
\begin{eqnarray}
  \vert k\rangle_x=\frac{1}{\sqrt{d}}\sum_{j=0}^{d-1}\exp(\frac{2\pi i}{d}jk)\vert j\rangle.
\end{eqnarray}
If any one of the $M$ controllers does not participate, the density matrix of the receiver's state is
\begin{eqnarray}
  \rho_B=\frac{1}{d}\vert \varphi\rangle\langle\varphi\vert+\frac{1}{d}\sum_{j=1}^{d-1}\vert \phi_j\rangle\langle\phi_j\vert.
\end{eqnarray}
The NCF is $\bar{f}=\frac{2}{d+1}$ which is exactly the classical limit. We can conclude that this scheme is a suitable one for CRSP since it preserves the controller's authority. The authors also discussed the CRSP of $N$-qudit quantum states via $N$ $(M+2)$-particle generalized GHZ-class states. Each controller holds $N$ particles. In this case, without any one controller's help, the receiver's state is
\begin{eqnarray}
\rho_B= \frac{1}{d^N}\vert \varphi\rangle\langle\varphi\vert+\frac{1}{d^N}\sum_{j=1}^{d^N-1}\vert \phi_j\rangle\langle\phi_j\vert.
\end{eqnarray}
Then the control power is $\frac{d^N-1}{d^N+1}$, which confirms the suitability of this scheme for the CRSP task. These protocols use POVM measurements in addition to the partially entangled channel to achieve enough control power.

In Ref.\cite{CRSP_qudit2}, CRSP of a two-qudit state was proposed. The difference from  Ref.\cite{CRSP_qudit1} is that the quantum channel is composed of three generalized Bell states $\vert \Theta\rangle=\vert \Phi_{00}\rangle_{A_1B_1}\otimes\vert \Phi_{00}\rangle_{A_2C_1}\otimes\vert \Phi_{00}\rangle_{B_2C_2}$, where
\begin{eqnarray}
  \vert \Phi_{kl}\rangle=\frac{1}{\sqrt{d}}\sum_{j=1}^{d-1}\exp(\frac{2\pi i}{d}jk)\vert j\rangle\vert j\oplus l\rangle.
  \end{eqnarray}
The state to be prepared can be written as
\begin{eqnarray}
  \vert \varphi\rangle=\sum_{j_1j_2=0}^{d-1}\alpha_{j_1,j_2}\vert j_1j_2\rangle.
\end{eqnarray}
In this scheme, both the sender and the controller perform generalized Bell state measurements.
The density matrix of the receiver's state without the controller's help is
\begin{eqnarray}
\rho_B= \frac{1}{d^2}\vert \varphi\rangle\langle\varphi\vert+\frac{1}{d^2}\sum_{j=1}^{d^2-1}\vert \phi_j\rangle\langle\phi_j\vert.
\end{eqnarray}
The average NCF is $\frac{2}{d^2+1}$ and the control power is exactly the lower bound required for the scheme to be acceptable for CRSP.
We have thus found that these two CRSP schemes for high-dimensional states satisfy our criterion from the controller's viewpoint. The above analysis not only evaluates these controlled-RSP schemes from the controller's perspective, but also conversely verifies that our measure is useful and practical.

\section{Discussion and summary}

In this paper, we have investigated control power in the context of controlled-RSP, which may succeed probabilistically. We have defined a quantitative criterion for a protocol to be suitable for controlled-RSP of an arbitrary dimensional system from the controller's viewpoint. In order to remotely prepare a $D$-dimensional quantum system, the controller's power should be no less than $(D-1)/(D+1)$, where $D=\prod d_j$ and $d_j$ $(j=1,2,...N)$ is the dimension of the $j$th particle/subsystem. This conclusion also holds for controlled teleportation (CT) \cite{CP1,CP2}.

We have analyzed the controller's power in controlled-RSP schemes via partially entangled channels. Generally lack of maximal entanglement in the quantum channel can reduce the success probability of the protocol. However, counterintuitively, the success probabilities of some proposed schemes can be 100\% and independent of the degree of entanglement of the quantum channels. But in this case, it is the control power that is affected.  Our analysis shows that these channels are not eligible for controlled-RSP of arbitrary quantum states since they do not guarantee the minimum required control power. Thus the lack of entanglement can affect control power in the scenarios where it does not affect the success probability.  Therefore, the control power should be evaluated together with the success probability when assessing controlled-RSP and CT schemes.

It is interesting to analyse the factors that affect the control power and the success probability of a scheme utilizing a given quantum channel. Let us  compare the three-qubit GHZ state $\vert \Phi^+_3\rangle_{ABC}$, the MS state $\vert \Theta_{MS_3}\rangle_{ABC}$ and the three-qubit partially entangled GHZ state $\vert PGHZ\rangle=(a\vert 000\rangle+b\vert 111\rangle)_{ABC}$ by calculating the von Neumann entropy of their subsystems. The results are shown in Table. I.
\begin{table}[h]
\caption{Subsystem entropies of  3-qubit quantum channels.}
\begin{tabular}{c|c|c|c }
\hline
$State $ &$S(\rho_A)$ &$S(\rho_B)$ &$S(\rho_C)$\\ \hline
$GHZ$ & $=1$ &$=1$&$=1$\\
$MS$ & $=1$ &$=1$&$<1$\\
$PGHZ$ & $<1$ &$<1$&$<1$\\
  \hline
\end{tabular}
\end{table}
As we know, the GHZ state can be used for perfect CT of an arbitrary single-qubit state and deterministic CRSP of a single-qubit equatorial state while preserving the controller's power \cite{CRSP_equ}. The MS state can also be used in CRSP and CT with unit success probability, but the control power is insufficient \cite{CRSP_partial2,gao}. On the other hand, the PGHZ state does not  yield unit success probability or sufficient control power. From the table we can deduce that the controller's power depends on the entropy of the particles held by the controller. In the simplest case, the controller's entropy should be 1 bit for CRSP or CT of an arbitrary single-qubit state.

In Ref.\cite{CP2}, we showed that for teleporting an $N$-qubit  state, each controller should have access to at least $N$ qubits to ensure the minimum control power. In fact, the key point is not the number of particles but the information content, which can be measured by the entropy $S=-\sum p_i\log_2 p_i$. Here $p_i$ is the probability of the $i$th possible result.
For controlled teleportation or remote preparation of an arbitrary $N$-qubit state, the controller's entropy must be at least $S=\log_2 2^N=N$ bits. Similarly, the controller's entropy should satisfy $S \geq log_2 D$ for controlled-RSP or CT of an arbitrary $D$-dimensional state.
This result can be used not only for controlled-RSP and CT schemes, but also for other controlled quantum communication schemes. That is, if the information transmitted is equal to $X$ bits or $X$ qubits, the controller's entropy should be at least $X$ bits to ensure his/her authority.

In conclusion, we have presented the first analysis of controller's power in controlled-RSP schemes. We have defined a measure of control power for CRSP schemes that can be applied to both deterministic and non-deterministic protocols. We generalized our analysis beyond CRSP of qubits and gave a new lower bound for the required control power in schemes for $D$-dimensional states. We have analyzed several existing controlled-RSP protocols and pointed out the lack of control power in some proposed schemes. We found that partially entangled channels are unsuitable for controlled-RSP of arbitrary quantum states. We also identified the requirement on the controller's entropy to ensure his/her authority. Moreover, our criteria are not only suitable for CT and controlled-RSP schemes but also applicable to other quantum communication protocols implemented in a controlled manner. Our results thus have broad relevance for designing robust communication protocols. We emphasize that the control power should be given the same importance as the success probability when evaluating any controlled quantum communication protocol.

\section*{Acknowledgement}

XL is supported by the National Natural Science
Foundation of China under Grant No. 11574038 and the Fundamental Research Funds for the Central Universities under
Grant No.CQDXWL-2012-014. SG acknowledges support from the Ontario Ministry of Research and Innovation and the Natural Sciences and Engineering Research Council of Canada.

\end{document}